\documentclass[aps,preprint,showpacs]{revtex4}
\usepackage{graphicx}
\usepackage{dcolumn}

\begin{document}
\preprint{ADP-11-21/T743}

\title{Production of a cascade hyperon in the K$^-$ - proton interaction} 
\author{R. Shyam$^{1,2}$, O. Scholten$^{3}$ and A.W. Thomas$^1$}
\affiliation {
$^1$Centre for the Subatomic Structure of Matter (CSSM),
School of Chemistry and Physics, University of Adelaide, SA 5005, Australia
 \\ 
$^2$ Saha Institute of Nuclear Physics, 1/AF Bidhan Nagar, Kolkata 700064, 
India  \\
$^3$ Kernfysisch Versneller Instituut, University of Groningen, NL-9747 AA
Groningen, The Netherlands}

\date{\today}

\begin{abstract}
We investigate the production of a cascade hyperon ($\Xi$) through the 
$K^- + p \to K^+ (K^0) + \Xi^- (\Xi^0)$ reactions, within an effective 
Lagrangian model where these reactions proceed via excitations of $\Lambda$ 
and $\Sigma$ resonance intermediate states in $s$- and $u$-channels. 
The coupling constants at the various vertices are taken from previous 
studies and SU(3) symmetry considerations. The calculated total cross 
sections of these reactions, which are in good agreement with the available 
data, are dominated by the contributions from the $\Lambda(1520)$ 
intermediate resonant state.  However, the $K^+$ meson angular distributions 
show selectivity to other resonant states in different angular regions and 
interference among these states leads to their strong backward peaking.
\end{abstract}
\maketitle
\newpage
Spectroscopy of hadrons is one of the key tools for studying quantum 
chromodynamics (QCD) in the non-perturbative regime. Lattice simulations, 
which provide the only {\it ab initio} calculations of QCD in this 
regime, are now able to reproduce a large part of the experimentally 
observed ground state hadron spectrum  (see, e.g.,~Refs.~\cite{dav04}). 
However, only a small subset of the excited state hadron spectrum is 
currently amenable to lattice calculations. For the success of this 
endeavor, it is highly desirable to have a large amount of data on the 
excited state spectrum, including in particular those hadrons where the 
widths associated with the states are not large - so that they can be 
easily identified~\cite{mah10}.

A major advantage of investigating the double strangeness ($S = -2$) 
$\Xi$ states is that they are much narrower than the $N^*$, $\Delta^*$, 
$\Lambda^*$ and $\Sigma^*$ states, which reduces the overlap complications 
with the neighboring states. Furthermore, two of the three valence quarks 
in the $\Xi$ are heavier than the third one, which cuts down the 
uncertainties in the extrapolations of lattice QCD calculations of the 
cascade masses~\cite{mat08}. This also makes them useful for the measurement 
of isospin symmetry breaking - in this case the interchangeability of an up 
and down quark~\cite{mil90}. There are only two cascade particles of any 
particular mass state with just this type of quark interchange: $\Xi^-$ and 
$\Xi^0$.  On the experimental side, the detached decay vertex for many 
cascades allows their easier separation from various backgrounds.   

In contrast to $S = -1$ hyperons, the $\Xi$ states are underexplored.
Out of more than twenty $\Xi$ candidates expected in the SU(3) multiplet 
and at least ten such candidates predicated by the quark model calculations 
of Ref.~\cite{cap86}, only two ground state cascades, $\Xi$ and $\Xi(1530)$, 
are established with near certainty, as indicated by their four star status
in the latest review published by the particle data group~\cite{pdg10}. 
The reason is that the cross sections of $S = -2$ hyperons are relatively 
small, with the bulk of the cascade production data having been collected 
by studies of the $K^- + p \to K^+ + \Xi^-$ and $K^- + p \to K^0 + \Xi^0$ 
reactions in the sixties and early seventies using the hydrogen bubble 
chambers~\cite{pje62,dau69}. The total cross section data from 
these measurements are tabulated in Ref.~\cite{fla83}. More recently, 
$\Xi^-$ production has been studied at Jefferson Laboratory via the reaction 
$\gamma p \to K^+K^+\Xi^-$, using photon beams with energies varying from 
2.75 GeV to 3.85 GeV~\cite{guo07}. In this experiment no significant 
signal of excited $\Xi$ states other than $\Xi(1530)$ has been observed.

Some attempts have been made in the past to understand the mechanism of the 
$K^- + p \to K^+ + \Xi^-$ reaction within simplified one meson and two meson
exchange models~\cite{ebe67,mir82}. Both approaches were unable to describe 
properly the existing data even though the two-meson exchange mechanism was 
somewhat better in this regard. The proper understanding of this reaction 
within a rigorous model is important for several reasons. A strong program 
is proposed at the JPARC facility in Japan and eventually at GSI-FAIR in 
Germany to obtain information about the spectroscopy of $\Xi-$hypernuclei 
through the $(K^-,K^+)$ reaction on nuclear targets using the high intensity 
and high momentum $K^-$ beam. Establishing the existence and properties of 
$\Xi$ hypernuclei is of considerable importance for a number of reasons
\cite{gui08}, not least as a constraint on the role of $\Xi$ hyperon in dense 
matter at the core of the neutron star. The $K^- + p \to K^+ + \Xi^-$ reaction
is the best tool to implant a $\Xi$ hyperon in the nucleus through the 
$(K^-,K^+)$ reaction. Coupled with recent progress in lattice QCD~\cite{bea11},
the availability of a high quality $K^-$ beam is likely to revive interest in 
looking for a near stable six-quark dibaryon resonance ($H$) with spin-parity 
of $0^+$, isospin 0 and $S=-2$~\cite{jaf77,mul83}, by studying the 
$(K^-,K^+)$ reaction. The amplitude for the $K^- + p \to K^+ + \Xi^-$ process
must be known accurately in order to estimate the cross section for  
$H$ production~\cite{aer82}.  

\begin{figure}
\centering
\includegraphics[width=.50\textwidth]{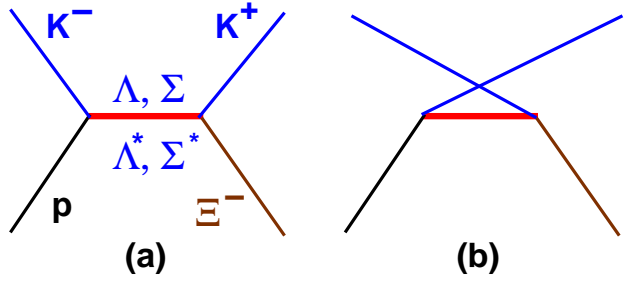}
\caption{[color online]
Graphical representation of our model to describe $K^- + p \to K^+ + \Xi^-$ 
reaction. 
}
\label{fig:Fig1}
\end{figure}
\noindent

In this paper we investigate the $K^- + p \to K^+ + \Xi^-$ and $K^- + p 
\to K^0 + \Xi^0$ reactions within a single channel effective Lagrangian 
model, which is similar to that developed in Ref.~\cite{shy09} to study 
the associated photoproduction of kaons off protons. These reactions are 
clean examples of a process in which baryon exchange plays the dominant 
role and the $t-$channel meson exchanges are absent, as no meson with 
$S=+2$ is known to exist. In our model, contributions are included from 
$s$- and $u$-channel diagrams (Figs.~1(a) and 1(b), respectively), which 
have as intermediate states $\Lambda$ and $\Sigma$ hyperons together with 
eight of their three and four star resonances with masses up to 2.0 GeV 
[ $\Lambda(1405)$, $\Lambda(1520)$, $\Lambda(1670)$, $\Lambda(1810)$, 
$\Lambda(1890)$, $\Sigma(1385)$, $\Sigma(1670)$ and $\Sigma(1750)$, which 
are represented by $\Lambda^*$ and $\Sigma^*$ in Fig.~1]. In past studies 
of these reactions~\cite{ebe67,mir82}, such a comprehensive investigation 
of the influence of so many intermediate resonance states has not been 
attempted. Moreover, authors of Refs.~\cite{ebe67} have not shown 
calculations for both hyperon production channels within their respective 
models. 

We would like to add that calculations of this reaction within a coupled 
channels model along the lines of those presented in Refs.
\cite{shy08,shy10,pen02}, are in principle, possible, where e.g., 
$n$ + $K^0$ intermediate states can also be included. Some contributions 
from the coupled channels effects with two sequential $K$ or $K^*$ exchanges 
and intermediate $\pi/\rho$ meson + $\Lambda/\Sigma$ states are also 
possible. Such calculations are under development.

The form of the effective Lagrangian vertices involving spin-$\frac{1}{2}$ 
resonance intermediate states are taken as (see, 
Ref.~\cite{shy08}) 
\begin{eqnarray}
L_{KBR_{1/2}} & = & -g_{KBR_{1/2}} [\chi\,{i\Gamma}\, {\varphi_K}+
                   \frac{(1-\chi)}{M}\,\Gamma\, \gamma_\mu\,(\partial^\mu 
                   \varphi_K)]   
\end{eqnarray} 
with $M \,=\,(m_R \,\pm\,m_B)$, where the upper sign corresponds to an 
even-parity and the lower sign to an odd-parity resonance, and B represents 
either a nucleon or a $\Xi$ hyperon. The operator $\Gamma$ is $\gamma_5$ (1) 
for an even- (odd-) parity resonance. The parameter $\chi$ controls the 
admixture of pseudoscalar and pseudovector components. The value of this 
parameter is taken to be 0.5 for the $\Lambda^*$ and $\Sigma^*$ states, but 
zero for $\Lambda$ and $\Sigma$ states implying pure pseudovector 
couplings for the corresponding vertices, which is in agreement with Refs.
\cite{shy08,shy99}. For spin-$\frac{3}{2}$ resonance vertices, 
we have used the gauge-invariant effective Lagrangian as given in Refs.
\cite{shy08}. The corresponding vertex function is written as
\begin{eqnarray}
L_{KBR_{3/2}}^\alpha &=& \frac{g_{KBR_{3/2}}}{m_K}\,
   \Big[ \gamma^\alpha \, (q \cdot p) -{p\!\!\!/} q^\alpha \Big]
    \Big[(1-\chi) + \chi\, {p\!\!\!/} /M_B \Big],
\end{eqnarray}
where $p$ is the four momentum of the resonance and $q$ is that of the 
meson. Index $\alpha$ belongs to the spin-$\frac{3}{2}$ spinor. An 
interesting property of this vertex is that the product 
$\gamma \cdot L = 0$. As a consequence the spin-$\frac{1}{2}$ part of the 
corresponding propagator becomes redundant and only the spin-$\frac{3}{2}$ 
part gives rise to nonvanishing matrix elements~\cite{kon00,pas00}. 

We have used the following form factor at various vertices in both $s$- and 
$u$-channels
\begin{eqnarray}
  F_m(s)=\frac{\lambda^4}{\lambda^4+(s-m^2)^2},
\end{eqnarray}
where $m$ is the mass of propagating particle and $\lambda$ is the cutoff
parameter which is taken to be 1.2 GeV everywhere. Isospin manipulations 
have been done separately, giving rise to additional constant factors at 
each vertex.
\squeezetable{
\begin{table}[here]
\caption[T1]{$\Lambda$ and $\Sigma$ resonance intermediate states included 
in the calculation. 
}
\begin{ruledtabular}
\begin{tabular}{|c|c|c|c|c|c|}
\hline
Intermediate state & $L_{IJ}$ &$M$      &${Width}$    &$g_{KRN}$ &$g_{KR\Xi}$\\
 ($R$)                  &          &(\footnotesize{GeV})&(\footnotesize{GeV})&  &
\\
\hline
$\Lambda$ &                   &1.116    &0.0         &-16.750  & 10.132  \\
$\Sigma $ &                   &1.189    &0.0         &  5.580  &-13.50   \\
$\Lambda(1405)$ &$ S_{01}$    &1.406    &0.050       &1.585    & -0.956  \\ 
$\Lambda(1670)$ &$ S_{01}$    &1.670    &0.035       &0.300    & -0.182  \\ 
$\Lambda(1810)$ &$ P_{01}$    &1.180    &0.150       &2.800    &  2.800  \\ 
$\Lambda(1890)$ &$ P_{03}$    &1.890    &0.100       &0.800    &  0.800  \\ 
$\Lambda(1520)$ &$ D_{03}$    &1.520    &0.016       &27.46    &-16.610  \\ 
$\Sigma (1750)$ &$ S_{11}$    &1.750    &0.090       &0.500    &  0.500  \\ 
$\Sigma (1385)$ &$ P_{13}$    &1.383    &0.036       &-6.22    & -6.220  \\ 
$\Sigma (1670)$ &$ D_{13}$    &1.670    &0.060       & 2.80    &  2.800  \\ 
\hline
\end{tabular}
\end{ruledtabular}
\end{table}
}
 
In Table I, we have listed the spin-parities, masses and widths of all the
intermediate resonance states included in our calculations. Also given
there are the coupling constants (CC) that have been used in our 
calculations at various vertices. For $K\Lambda N$ and $K\Sigma N$ vertices, 
the CC adopted by us are on the upper side of those obtained within the SU(3) 
model~\cite{sam74,dom83,gen04} with the $\alpha_D$ parameter (the standard 
fraction involving D and F couplings) of 0.644~\cite{don82} which is very
close to the SU(3) value. But given that SU(3) symmetry is broken at the 
level of 20$\%$ there may be uncertainty in these values of this order
\cite{ade90}.  
The relative signs of these couplings were fixed by the SU(3) predictions. For 
the corresponding CC involving the $\Xi$ hyperon, the SU(3) relations as 
given in Ref.~\cite{swa63} have been used.

On the other hand, the knowledge about the CC for $KRN$ vertices (where $R$ 
represents a $\Lambda^*$ or $\Sigma^*$ state) is very scanty. Even 
more pathetic is the situation regarding the CC of corresponding vertices 
involving the $\Xi$ hyperon where little or no information exists. 
In this study, the magnitudes of the CC for the $KRN$ vertices involving 
the low-lying resonances, $\Lambda(1405)$, $\Lambda(1520)$, and $\Sigma(1385)$,
have been adopted from those given in Ref.~\cite{bra77}, while those of the 
high energy resonances are determined from their decay widths as listed in 
Ref.~\cite{pdg10}. For the signs of these couplings, we were guided by the 
SU(3) predictions~\cite{sam74}, wherever possible. The CC of the $KR\Xi$ 
vertices are taken to be equal to those of the corresponding $KRN$ vertices
for high mass resonances, as suggested in Refs.~\cite{kin63,nak06}, whereas 
the SU(3) relations have been used to determine them for the lower mass 
hyperon states.   
\begin{figure}
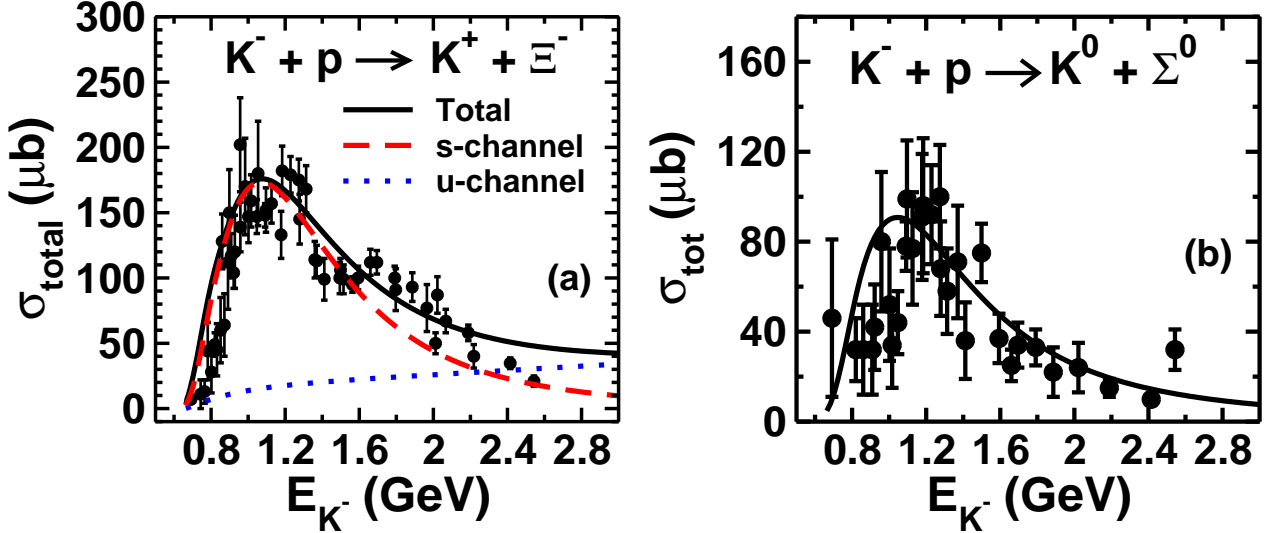

\begin{tabular}{cc}
\centering
\includegraphics[width=.50\textwidth]{Fig2a.eps}\hspace{0.20cm}
\includegraphics[width=.50\textwidth]{Fig2b.eps}
\end{tabular}
\caption{[color online]
Comparison of the calculated total cross section for the 
$K^- + p \to K^+ + \Xi^-$ (a) and $K^- + p \to K^0 + \Xi^0$ (b) reactions 
as a function of incident $K^-$ kinetic energy with the corresponding 
experimental data from Ref.~\protect\cite{fla83}. Also shown are the 
individual contributions of $s$- and $u$-channel diagrams to the total 
cross section for the reaction shown in panel (a).  
}
\label{fig:Fig2}
\end{figure}
 
In Fig.~2, we show comparisons of our calculations with the data for
the total cross sections of the $K^- + p \to K^+ + \Xi^-$ (panel a) and 
$K^- + p \to K^0 + \Xi^0$ (panel b) reactions for $K^-$ beam energies 
($E_{K^-}$) below 3.0 GeV because the resonance picture is not suitable at 
energies higher than this. It is clear that our model is able to reproduce 
the data well for both the channels within the statistical errors. We shall 
mostly be discussing the $K^- + p \to K^+ + \Xi^-$ reaction in the rest of 
this paper. In this case both calculated and experimental cross sections 
peak at $E_{K^-}$ $\approx 1.1$ GeV. We further note that the cross sections 
around the peak and the tail ($E_{K^-} \geq 2.1$ GeV) regions are dominated 
by the $s$- and $u$-channel contributions, respectively. This result is in 
contrast to the conclusions of past studies~\cite{ebe67,kin63} where 
$u$-channel contributions dominated this reaction everywhere.    
\begin{figure}
\centering
\includegraphics[width=.50\textwidth]{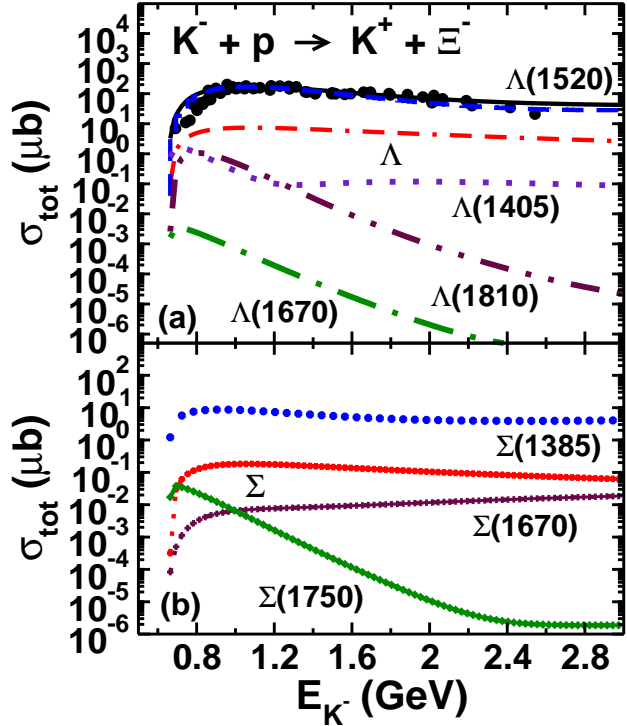}
\caption{[color online]
Contributions of individual $\Lambda$ and $\Lambda^*$ [panel (a)] and 
$\Sigma$ and $\Sigma^*$ [panel (b)] resonance intermediate states to the 
calculated total cross section for the same reaction as that shown in 
Fig.~2.  
}
\label{fig:Fig3}
\end{figure}

From Fig.~3, we note that the contribution from the $\Lambda(1520)$ 
intermediate state dominates the total cross sections over the entire regime
of $E_{K^-}$ values. We have checked that both $s$- and $u$-channel cross 
sections are also individually dominated by this resonance. The $\Lambda$, 
$\Lambda(1405)$, $\Sigma(1385)$ states make noticeable contributions only 
for $E_{K^-}$ very close to the production threshold. Other resonances 
contribute very weakly. Of course, our results are quite dependent of the 
CC of various vertices which are currently quite uncertain. Nevertheless, 
the relative cross sections shown in Fig.~3 are robust despite this. There 
is very little scope for increasing further the individual contributions of 
the $\Lambda$ and $\Sigma$ intermediate states, because the CC of the 
corresponding vertices used by us are already larger than the upper limits 
for them suggested in the literature (as discussed above). Furthermore, 
except for the $\Sigma(1385)$ resonance, where we have again used a 
larger CC,  the contributions of other resonances are too weak and even have 
the wrong $E_{K^-}$ dependence. Therefore, the final results are unlikely 
to be affected too much by the known uncertainties in the corresponding CC.
 
It is interesting to note that in a recent study of this reaction
~\cite{sha11}, where the data were fitted by a phenomenological model, the 
inclusion of $\Lambda(1520)$ and $\Sigma(1385)$ resonances improved the 
fits considerably. Nevertheless, the quality of the agreement with the 
total cross section data obtained by these authors is considerably poorer
than those shown in Figs. 2(a) and 2(b) if resonances with masses below 2 
GeV only are included. Moreover, in this reference neither the contributions 
of individual resonances nor the coupling constants at various vertices 
are shown explicitly.
\begin{figure}
\centering
\includegraphics[width=.50\textwidth]{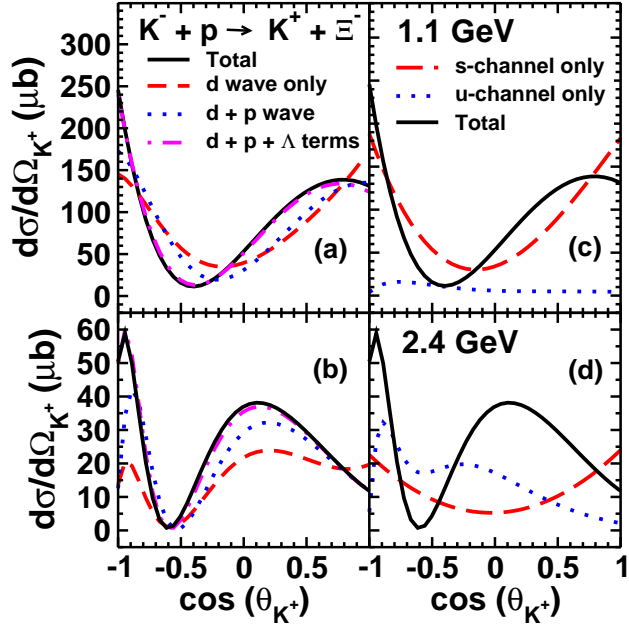}
\caption{[color online]
Contributions of resonance intermediate states of various partial waves 
[(a) and (b)], and of $s$- and $u$-channel terms [(c) and (d)] to the 
angular distributions of $K^+$ meson at beam energies of 1.1 GeV and 2.4 GeV,
respectively.  
}
\label{fig:Fig4}
\end{figure}

In Fig.~4, we have shown results for the angular distribution of the 
$K^+$ meson at $E_{K^-}$ values of 1.1 GeV (where the total cross section 
peaks) and 2.4 GeV (where $u$-channel contributions dominate these cross
sections). Differential cross sections provide more valuable information 
about the reaction mechanism because they include terms that weigh the 
interference terms of various components of the amplitude with outgoing 
$K^+$ angles. From panels (a) and (b), we notice that although the 
contributions of $p$-wave resonances and the $\Lambda$ terms are rather 
small for the total cross sections, their interference with the dominant 
$d$-wave $[\Lambda (1520) D_{03}]$ resonance influences the angular 
distributions strongly at both beam energies. This leads to enhanced 
backward peaking of the cross sections at both energies, and causes the 
forward bending of the angular distributions at 1.1 GeV and the increase in 
the forward peak value of the cross sections at 2.4 GeV. The magnitude
of the interference effects are directly related to those of the individual
intermediate resonance states. Therefore, the proper knowledge of the 
corresponding coupling constants is vital in this respect. Hence the 
angular distributions data are expected to put limits on the vertex
constants of even those resonances that contribute only weakly to the
total cross sections. The same is true also for the polarization data.

The role of interference terms of the $s$- and $u$-channel amplitudes
is studied in panels (c) and (d) of this figure. We note that here too
the interference terms are significant - it is somewhat of lesser importance
at $E_{K^-} = $ 1.1 GeV beam energy but is quite strong at 2.4 GeV. Therefore
the $s$-channel resonance contributions cannot be ignored in the description
of the angular distributions of the $K^- + p \to K^+ + \Xi^-$ reaction,
even at higher beam energies.

We would like to remark that some data on the differential cross sections 
of the $K^- + p \to K^+ + \Xi^-$ reaction are presented in Ref.~\cite{dau69}.
We, however, found it difficult to use this data due to several reasons. 
Firstly, it is not clear what the data actually represent - whether they 
are the angular distributions of the c.m. of the final channel or those of 
the $\Lambda$ hyperon. Secondly, the data have unknown normalizations and 
corrections; the integration over angles do not lead to the total 
cross sections reported by the same authors. Finally, in the absence of any 
table giving the data in the numerical form, it is not easy to reproduce them 
correctly with error bars. On the other hand, because the angular 
distributions of the Kaons are most likely to be measured at the JPARC 
facility, we presented our calculations for this case. 

In summary, we studied the $K^- + p \to K^+ + \Xi^-$ reaction within an
effective Lagrangian model that has $s$- and $u$-channel diagrams involving
as intermediate states the $\Lambda$ and $\Sigma$ hyperons together with 
their eight resonance states having masses below 2.0 GeV. The magnitudes and 
signs of the coupling constants at various vertices have been chosen from 
those determined in previous studies and from the prediction of the SU(3) 
model. 

An important result of our study is that total cross section of this 
reaction is dominated by the contributions from the $\Lambda(1520)$ 
(with $L_{IJ}=D_{03}$) resonance intermediate state through both $s$- and 
$u$-channel terms. The peak region gets most contributions from the 
$s$-channel graphs, while in the tail region ($E_{K^-} > 2.0$ GeV) the 
$u$-channel terms are dominant. However, the angular distributions of the 
outgoing $K^+$ meson show sensitivity to those resonance states that 
otherwise contribute only weakly to the total cross section. In particular,
the strong backward peaking of the $K^+$ differential cross sections results 
from the interference effects of various intermediate states in both $s$- 
and $u$-channel terms. This result is of vital significance as it invalidates
the long standing belief that the strong backward peaking of the angular 
distributions in this reaction results from the $u$-channel dominance. Of 
course, $u$-channel diagrams are important particularly at higher beam 
energies. However, consideration of the s-channel resonances is important 
at these energies too. We stress that measurements of the outgoing kaon 
angular distributions and polarizations are of crucial importance for putting 
constraints on largely unknown vertex parameters for the decay of 
$\Lambda$ and $\Sigma$ resonances to the kaon-baryon channels. This 
study is a precursor to the theory of the production of $\Xi$-hypernuclei 
via $(K^-,K^+)$ reactions which is being developed. 

This work has been supported by the University of Adelaide and the Australian 
Research Council through grant FL0992247(AWT).

\end{document}